\documentclass[prl,final,twocolumn,superscriptaddress,showpacs]{revtex4}
\usepackage{graphicx}

\begin{document}

\title{From Quantum criticality to enhanced thermopower in strongly correlated layered cobalt oxide}

\author{P. Limelette$^\dagger$}
\author{W. Saulquin $^\ddagger$}
\affiliation{Universit\'e Fran\c{c}ois Rabelais, $^\dagger$Laboratoire LEMA, UMR 6157 CNRS-CEA, $^\ddagger$D\'epartement de Physique, Parc de Grandmont, 37200 Tours, France}
\author{H. Muguerra}
\affiliation{Inorganic Materials Chemistry, Dept. of Chemistry, University of Li\`ege, All\'ee de la Chimie 3 (Bat. B6), 4000 Li\`ege, Belgium}
\author{D. Grebille*}
\affiliation{Laboratoire CRISMAT, UMR 6508 CNRS-ENSICAEN et Universit\'e de Caen, 6, Boulevard du Mar\'echal Juin, 14050 CAEN Cedex, France}

\begin{abstract}
\vspace{0.3cm}
We report on susceptibility measurements in the strongly correlated layered cobalt oxide [BiBa$_{0.66}$K$_{0.36}$O$_2$]CoO$_2$ which demonstrate the existence of a magnetic QCP governing the electronic properties. The investigated low frequency susceptibility displays a scaling behaviour with both the temperature T and the magnetic field B ranging from the high-T non-Fermi liquid down to the low-T Fermi liquid. Whereas the inferred scaling form can be discussed within the standard framework of the quantum critical phenomena, the determined critical exponents suggest an unconventional magnetic QCP of a potentially generic type. Accordingly, these quantum critical fluctuations account for the anomalous logarithmic temperature dependence of the thermopower. This result allows us to conjecture that quantum criticality can be an efficient source of enhanced thermopower.
\end{abstract}

\pacs{71.27.+a}

\maketitle

The strongly correlated quantum matter exhibits various outstanding and often puzzling properties related to quantum criticality \cite{Gegenwart2008}.
Providing a route towards non-Fermi Liquid behaviour or unconventional superconductivity, these quantum fluctuations originate from a so-called quantum critical point (QCP).
Located at zero temperature, a QCP results from competing interactions which can be tuned by an appropriate non-thermal control parameter, namely such as pressure, doping or magnetic field \cite{Sachdev1999}.
While heavy Fermions metals have early emerged as prototypical materials to investigate QCPs \cite{Custers2003,L¨ohneysen2007,Schr¨oder2000}, some transition-metal oxides as the ruthenates \cite{Gegenwart2006} and the cuprate high-Tc superconductors \cite{VanderMarel2003} have also revealed amazing properties related to quantum criticality.
In addition, most of these oxides share in common that they are doped Mott insulator, i.e. their metallicity originates from the introduction of charge carriers by doping, otherwise the strong Coulomb repulsion would localize electrons to form a Mott insulating state \cite{Georges1996}.

Belonging to this class of materials, the layered cobalt oxides have revealed, besides their enhanced room temperature thermopower \cite{Terasaki1997,Limelette2006}, striking properties \cite{Bobroff2007} including superconductivity \cite{Takada2003}, large negative magnetoresistance in some compounds \cite{Limelette2008}, or giant electron-electron scattering in Na$_{0.7}$CoO$_2$ \cite{Li2004}.
The latter observation has already led to conjecture a possible influence of a magnetic QCP.
Interestingly, density functional calculations \cite{Singh2007} have predicted at the local spin-density approximation level weak itinerant ferromagnetic state competing with weak itinerant antiferromagnetic state in electron doped Na$_x$CoO$_2$.
Due to these competing interactions, quantum critical fluctuations have been speculated and even a possible triplet superconductivity. 
Here we report on the low frequency susceptibility measurements which allow us to identify a magnetic quantum critical point in the layered cobalt oxide [BiBa$_{0.66}$K$_{0.36}$O$_2$]CoO$_2$ leading to anomalous thermoelectric behaviour.

\begin{figure}[h]
\centerline{\includegraphics[width=0.9\hsize]{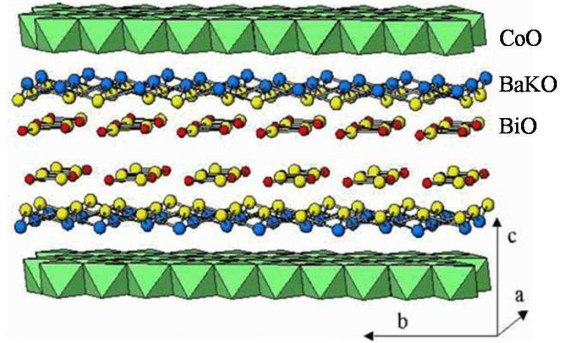}}
\caption{(Color online). Schematic picture of the crystal structure of the layered cobalt oxide [BiBa$_{0.66}$K$_{0.36}$O$_2$]CoO$_2$.  It displays from the top to the bottom a CoO$_2$ plane, a Ba$_{0.66}$K$_{0.36}$O plane, a BiO plane followed by symmetrical layers.
}
\label{fig0}
\end{figure}

\begin{figure}[h]
\centerline{\includegraphics[width=0.95\hsize]{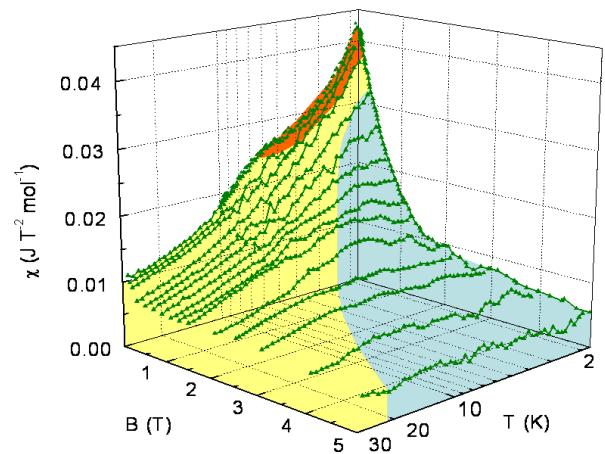}}
\caption{(Color online). Temperature dependence of the low frequency susceptibility $\chi$ as a function of the out of plane magnetic field B.
The areas show three distinct behaviours emerging from a magnetic QCP near the critical field B$_C$=0.176 T.
The dark gray (red) area displays the gradual destruction of a magnetically ordered state with the disappearance of the anomaly observed around 6 K. $\chi$ recovers a Fermi liquid Pauli-like behaviour at higher magnetic fields over an extended temperature range as B increases.
This results from a magnetic field dependent cross-over from a low-T Fermi liquid in gray (blue area) up to a high-T non-Fermi liquid in light gray (yellow area).
}
\label{fig1}
\end{figure}

\begin{figure}[htbp]
\centerline{\includegraphics[width=0.88\hsize]{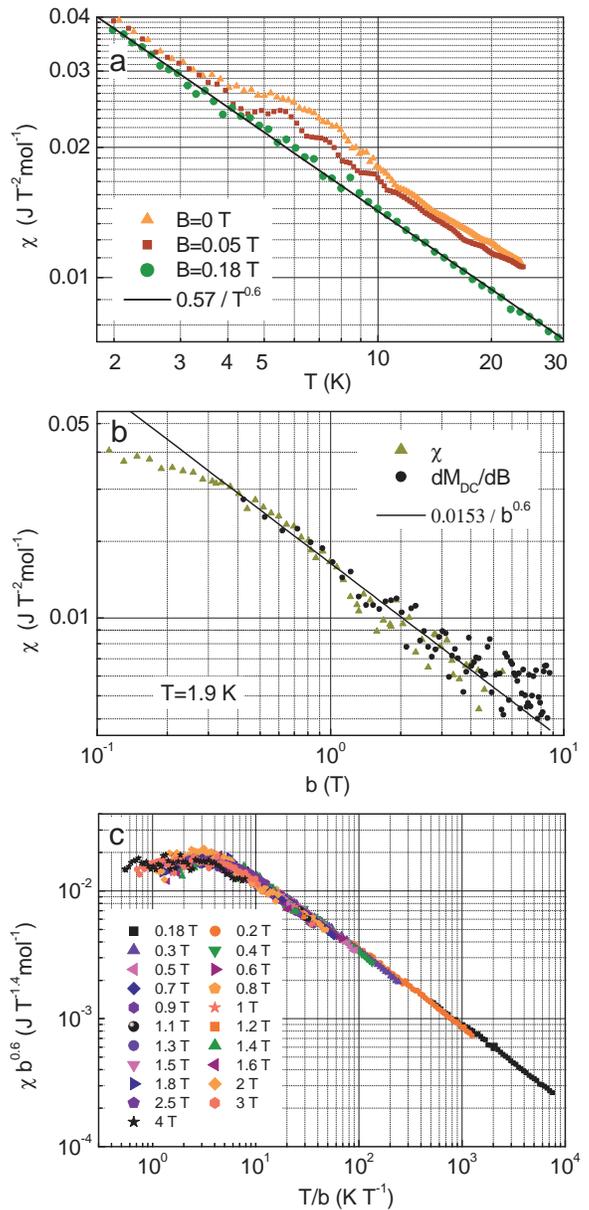}}
\caption{(Color online). \textbf{a}: gradual disappearance of the magnetic anomaly around 6 K and power law T-dependence of the non-Fermi liquid susceptibility in the vicinity of B$_C$=0.176 T.
\textbf{b}: susceptibility measured at T=1.9 K showing a power law dependence as a function of the magnetic distance b=B-B$_C$ from the QCP.
A comparison is here made between the $\chi$ and the differential susceptibility defined as dM$_{DC}$/dB and discussed in the text.
\textbf{c}: scaling plot of the susceptibility evidencing that the whole data agrees with both the fractional exponents determined in a and b within the range B$>$B$_C$.
}
\label{fig2}
\end{figure}

Similarly to Na$_x$CoO$_2$, the structure of the layered cobalt oxide [BiBa$_{0.66}$K$_{0.36}$O$_2$]CoO$_2$ contains single CoO$_2$ layer of CdI$_2$ type stacked with four rocksalt-type layers as depicted in Fig. \ref{fig0}, instead of a sodium deficient layer, which act as a charge reservoir \cite{Hervieu2003}.
The reported measurements have been performed on single-crystals, which were grown using standard flux method \cite{Leligny1999}.
The low frequency susceptibility has been determined with an ac field modulation up to 1.7 mT at a frequency of 1 kHz superposed to a constant field B using an AC-magnetometer of a Quantum Design physical properties measurement system.
A stacking of few single-crystals, all oriented with c axis parallel to the magnetic field, has been used to reach a mass of 20 mg for these measurements.
Figure \ref{fig1} displays the temperature dependences of the susceptibility for selected values of magnetic field which basically span three regimes with specific behaviours as highlighted by the areas.
In particular, a magnetic field induced Fermi liquid regime characterized by a temperature independent susceptibility is revealed at the lowest temperatures (blue or gray area).
This behaviour is followed by a power law T-dependence at higher temperatures (yellow or light gray area).
At very low fields, an anomaly can be seen around 6 K indicating a possible enhancement of the magnetic correlations.
These anomalies disappear above nearly 0.2 T as exemplified in Fig.~\ref{fig2}a by delimitating the third regime in Fig.~\ref{fig1} (red or dark gray area).
Contrasting with the divergent behaviour of the susceptibility expected in the case of ferromagnetic correlations, such an anomaly likely suggests an antiferromagnetic ordering of spin density wave type for instance, generically labelled hereafter magnetically ordered state. The gradual destruction of the latter state with magnetic field concomitantly to the stabilization of the field induced Fermi liquid provides therefore the key ingredients for the occurrence of a magnetic QCP at the critical magnetic field B$_C \approx$ 0.176 T in Fig.~\ref{fig1}.
It is worth mentioning that the low magnetic field anomalies could also arise from disorder.
Nevertheless, such a scenario doesn't seem able to account for both the inferred universal power laws and the observed scaling behavior as discussed thereafter.

Within the Fermi liquid regime above B$_C$, the Pauli-like susceptibility which is temperature independent is found to exhibit a strong increase near the QCP.
This behaviour may originate from several scenarios.
Indeed, it could result from either an enhancement of the electronic effective mass due to the vicinity of the QCP \cite{Custers2003} or because of the presence of efficient ferromagnetic fluctuations increasing the Pauli susceptibility by a Stoner factor.
Of course, one cannot rule out a more complex situation involving the conjunction of both of these effects as quite recently observed in the case of the heavy Fermion compound YbRh$_2$(Si$_{0.95}$Ge$_{0.05}$)$_2$ \cite{Gegenwart2005}.

Furthermore, the magnetic field induced Fermi liquid regime is found to be restricted in Fig.~\ref{fig1} over a temperature range which continuously decreases by approaching the QCP and defines a cross-over up to a high-T non-Fermi liquid.
One must here emphasize the amazingly vast temperature interval associated to this cross-over which differs from the ones observed in the majority of the heavy Fermions compounds such as YbRh$_2$(Si$_{0.95}$Ge$_{0.05}$)$_2$ from more than one order of magnitude \cite{L¨ohneysen2007,Gegenwart2005}.
This result points out the strength of the quantum critical fluctuations in this strongly correlated oxide.

Contrasting with the aforementioned Pauli susceptibility, $\chi$ displays within the quantum critical regime an anomalous temperature dependence characterized in Fig.~\ref{fig2}a by a fractional exponent such as $\chi \propto$ T$^{-0.6}$.
In addition, the Pauli susceptibility reveals a power law magnetic field dependence displayed in Fig.~\ref{fig2}b such as $\chi \propto$ b$^{-0.6}$, with the field b=B-B$_C$ which measures the distance from the QCP.
In order to check the value of the previous exponent, the DC magnetization has been measured as a function of the magnetic field at 1.9 K.
It has allowed to infer a differential susceptibility dM$_{DC}$/dB in Fig.~\ref{fig2}b in a quantitative agreement with the low frequency susceptibility.
We also emphasize that these power laws are well defined over more than one decade ensuring thus the qualities of the fits. Interestingly, such fractional exponents have already been deduced in heavy fermion compounds near a magnetic QCP \cite{Custers2003,Schr¨oder2000,Gegenwart2005}.
In particular, exponents equals to 0.75 have been determined in the heavy Fermion metal CeCu$_{6-x}$Au$_x$ based on neutron scattering and bulk magnetometry measurements \cite{Schr¨oder2000}.
Here, the determined fractional exponents 0.6 are found to remarkably agree with the ones deduced more recently in the heavy Fermion metal YbRh$_2$(Si$_{0.95}$Ge$_{0.05}$)$_2$ near a magnetic QCP \cite{Gegenwart2005}.
Because of the differences between the heavy Fermion compounds and the layered cobalt oxides, these power law dependences could appear as intrinsic features of such a magnetic QCP, suggesting thus, fractional exponents potentially generic.

As a strong check of the previous power laws, Fig.~\ref{fig2}c demonstrates that the whole data set including both the Fermi and the non-Fermi liquid regimes (B$>$B$_C$) can be scaled over several decades onto an equation such as $\chi$(b,T)=b$^{-0.6} \varphi$(T/b), 
with b which measures the distance from the QCP.
The scaling function $\varphi$(T/b) is such that when T $\rightarrow$ 0, $\chi$ is expected to recover a temperature independent form according the Pauli susceptibility measured in the Fermi liquid regime.
It follows that $\varphi$(T/b $\rightarrow$ 0)  =  constant and thus, $\chi(b,T\rightarrow 0)\propto$ b$^{-0.6}$.
On the other hand, the susceptibility cannot depend on the field in the quantum critical regime when b = 0.
It results that $\varphi$(T/b$\rightarrow \infty$) $\propto$ (T/b)$^{-0.6}$ which leads to $\chi(b,T\rightarrow 0)\propto$ T$^{-0.6}$.

In order to discuss such a scaling form, let us now briefly introduce some basic ideas concerning the critical behaviour near a QCP.
In contrast to classical phase transitions at T$>$0, phase transitions at T=0 are dominated by quantum effects.
According to the modern picture introduced by Hertz and Millis \cite{Hertz1976}, these effects originate from an intrinsic dynamics traducing the scaling of the excitations frequency with wave vector such as $\omega \propto$ k$^z$, z being the so-called dynamical exponent \cite{L¨ohneysen2007}.
It follows that both the correlation length $\xi \propto$ b$^{-\nu}$ and the correlation time $\tau \propto$ $\xi^z$ are expected to diverge at the transition.
The effective dimensionality is then raised from the spatial dimension d by z such as d$_{eff}$=d+z, reflecting the mixing of statics and quantum dynamics.
Also, a transition with d$_{eff}$ larger than the upper critical dimension, namely d$_{eff} \geq$ d$^+$ with here d$^+$=4, falls into mean field descriptions.
If d$_{eff} <$ d$^+$, the phase transition is instead controlled by interacting fixed point and usually obeys strong hyperscaling properties \cite{L¨ohneysen2007}.
Within the latter context, the critical contribution to the free energy density should follow an homogeneity law as a function of the arbitrary scale factor $\lambda$ as f(b,T)=$\lambda^{-(d+z)}$ f($\lambda^{1/\nu}$b,$\lambda^{z}$T).
By choosing $\lambda$ = $\xi$, the scaling form f(b,T)=$\xi^{-(d+z)}$ $\Phi$($\xi^{z}$T) can be inferred, with the scaling function $\Phi$.
This leads to f(b,T)=b$^{\nu (d+z)} \Phi$(T/b$^{\nu z}$).

Accordingly, the critical contributions to the susceptibility can be deduced such as $\chi$ = $\partial^2$f/$\partial$b$^2$:
\begin{equation}
\chi(b,T)=b^{-(2-\nu (d+z))} \phi \left(\frac{T}{b^{\nu z}}\right)
\label{eq1}
\end{equation}
In order to describe a cross-over from a low-T Fermi liquid regime up to a high-T non-Fermi liquid regime, equation \ref{eq1} must obey the proper asymptotic behaviours.
In the former regime where T $\rightarrow$ 0, the susceptibility recovers a Pauli contribution, such as:
\begin{equation}
\chi(b,T \rightarrow 0)\propto b^{-(2-\nu (d+z))}
\label{eq2}
\end{equation}
This results from the limit $\phi$(T/b$^{\nu z} \rightarrow$ 0)= constant.
On the other hand, the quantum critical fluctuations dominate in the non-Fermi liquid regime and the susceptibility acquires a non vanishing temperature dependence.
Since this regime is in particular achieved in the limit b = 0, the resulting susceptibility cannot depend on the field.
By considering the asymptotic behaviour of the scaling function $\phi$(T/b$^{\nu z} \rightarrow \infty$)$\rightarrow$ (T/b$^{\nu z}$)$^{-(2-\nu (d+z))/\nu z}$, $\chi$ can be written as it follows.
\begin{equation}
\chi(b = 0,T)\propto T^{-(2-\nu (d+z))/\nu z}
\label{eq3}
\end{equation}
Interestingly, the comparison between equations \ref{eq2} and \ref{eq3} with the experimental power laws reported in Fig.~\ref{fig2}a and \ref{fig2}b yields to a partial determination of the critical exponents.
The combination of the two fractional exponents leads to $\nu$z=1 and then $\nu$d=0.4.
In order to fulfil the aforementioned hyperscaling hypothesis d$_{eff} <$ d$^+$, these results imply $\nu >$0.35 which is expected to be always satisfied since the lowest value is the mean field one $\nu$=1/2.
By using the latter mean field value, it must be emphasized that the product $\nu$z=1 is thus consistent with the dynamical exponent z=2, which is predicted if the magnetically ordered state is a spin density wave \cite{L¨ohneysen2007,Hertz1976}.
However, such a scenario leads to a dimensionality slightly lower than 1 by making the use of the product $\nu$d=0.4.
So, even if the hyperscaling scenario provides some insight into the power law dependences of the susceptibility as well as its scaling behaviour, it raises the question of the meaning of these puzzling exponents as well as the nature of the quantum criticality.

Beyond the quite naive hyperscaling hypothesis, the determined fractional exponents could otherwise be a signature of local fluctuations leading to a so-called locally critical point \cite{Si2001}.
Such a QCP is indeed characterized by the coexistence of long wavelength and spatially local critical modes.
This yields to anomalous both frequency and temperature dependences of the susceptibility with fractional exponents not only at the ordering wave vector but essentially everywhere else in the Brillouin zone.
Based on a microscopic analysis of the Kondo lattice model carried out within an extended dynamical-mean-field approach \cite{Si2001}, it has been shown that such an exponent, so-called $\alpha$ could be written as a function of both the Kondo coupling $\Lambda$ and the RKKY density of states $\rho_I$(I$_Q$) as $\alpha$=1/(2$\Lambda \rho_I$(I$_Q$).
Due to the competition between the Kondo and RKKY interactions, the product $\Lambda \rho_I$(I$_Q$) is then expected to be close to unity near the QCP leading thus to an exponent $\alpha$ not too far from 1/2.
Because this kind of QCP is expected if the spin fluctuations are quasi two dimensional, the studied layered cobalt oxide could therefore appear as a very promising candidate in order to characterize the local criticality in a strongly correlated metal.

On the other hand, it is worth noting that a phenomenological spin fluctuation theory for quantum tricritical point \cite{Misawa2009} has very recently succeeded in predicting such a fractional exponent.
According to this approach, a first order phase transition changes to a continuous one at zero temperature leading to a quantum tricriticality which seems able to explain several properties observed in some of the heavy fermion compounds near a QCP.
These challenging scenarios call now for further experimental investigations in order to discriminate whether this unconventional magnetic QCP is locally critical or altered by a quantum tricriticality.

\begin{figure}[htbp]
\centerline{\includegraphics[width=0.95\hsize]{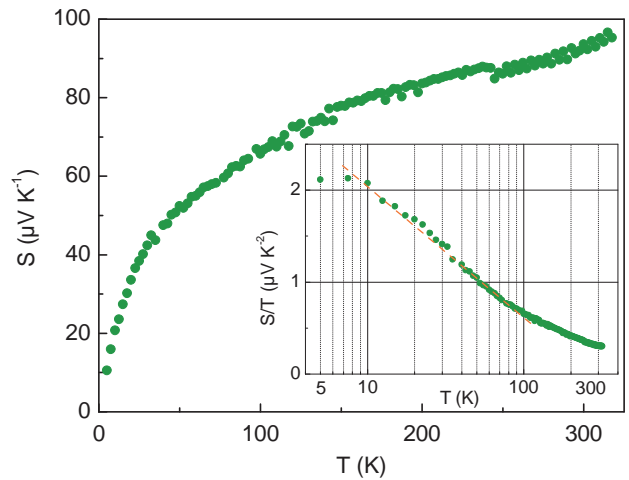}}
\caption{(Color online). Anomalous temperature dependence of the thermopower.
The inset displays S/T as a function of the temperature on a logarithmic scale.
The dash line is a guide to the eyes evidencing the quantum critical behaviour 
such as S/T $\propto$ ln(1/T) within the range from 10 K up to 100 K.
}
\label{fig3}
\end{figure}

Finally, let us briefly discuss the implications of such a quantum criticality within the context of the enhanced thermoelectric properties measured in these layered cobalt oxides \cite{Terasaki1997}.
As reported in Fig.~\ref{fig3}, the thermopower S displays at room temperature an unusually large value for a metal.
In addition, its T-dependence mainly contrasts with the linear variation expected in a metallic system \cite{Limelette2006}.

By plotting S/T as a function of the temperature on a logarithmic scale in the inset in Fig.~\ref{fig3}, two regimes can be inferred in a complete agreement with the overall reported analysis. 
Below nearly 10 K, the thermopower varies linearly with temperature with an enhanced coefficient such as roughly S $\approx$ 2.1 10$^{-6}$T VK$^{-1}$. 
Such a high coefficient likely originates from a very small number of delocalized charge carriers in accordance with an itinerant antiferromagnetic state as assumed from Fig. \ref{fig1}.
Indeed, due to the crystal symmetry, spin density wave type instability would not open a gap over the entire Fermi surface, keeping thus small pockets of delocalized charge carriers.
The latter state being destroyed by the quantum critical fluctuations above nearly 10 K, the variation of S/T within this new regime  turns out to decrease linearly over almost one decade indicating then a dependence as S/T $\approx$ 0.62 ln(260/T).
Such an anomalous thermoelectric behaviour, namely as S $\propto$ T ln(1/T), is a typical signature of a non-Fermi liquid state and it has been theoretically ascribed to the proximity of an antiferromagnetic QCP in a system where three-dimensional electrons are coupled to two-dimensional spin fluctuations \cite{Paul2001}.

Therefore, the electrons with momentum near the ordering wave vector undergo a singular scattering with the spin fluctuations, and a finite area of the Fermi surface will become hot.
It follows that the hot electrons are expected to carry an entropy larger than usual which leads to an increased thermopower with a logarithmic T-dependence.
It is worthwhile to notice that such a behaviour is observed in Fig.~\ref{fig3} over an amazingly wide range of temperatures up to nearly 100 K.
This result allows us to speculate that quantum critical fluctuations can provide an additional \cite{Wang2003} and efficient source of enhanced thermopower.
\begin{acknowledgments}
We thank Wataru Kobayashi for providing thermopower data and Sylvie H\'ebert for usefull discussions.
\end{acknowledgments}
\noindent * Deceased in January 2009.


\begin{thebibliography}{20}
\expandafter\ifx\csname natexlab\endcsname\relax\def\natexlab#1{#1}\fi
\expandafter\ifx\csname bibnamefont\endcsname\relax
  \def\bibnamefont#1{#1}\fi
\expandafter\ifx\csname bibfnamefont\endcsname\relax
  \def\bibfnamefont#1{#1}\fi
\expandafter\ifx\csname citenamefont\endcsname\relax
  \def\citenamefont#1{#1}\fi
\expandafter\ifx\csname url\endcsname\relax
  \def\url#1{\texttt{#1}}\fi
\expandafter\ifx\csname urlprefix\endcsname\relax\def\urlprefix{URL }\fi
\providecommand{\bibinfo}[2]{#2}
\providecommand{\eprint}[2][]{\url{#2}}

\bibitem[{\citenamefont{Gegenwart et~al.}(2008)\citenamefont{Gegenwart}}]{Gegenwart2008}
\bibinfo{author}{\bibfnamefont{P.}~\bibnamefont{Gegenwart}},
  \bibinfo{author}{\bibfnamefont{Q.}~\bibnamefont{Si}} \bibnamefont{and}
  \bibinfo{author}{\bibfnamefont{F.}~\bibnamefont{Steglich}}, 
  \bibinfo{journal}{Nature Physics} \textbf{\bibinfo{volume}{4}},
  \bibinfo{pages}{186} (\bibinfo{year}{2008}).
  


\bibitem[{\citenamefont{Sachdev}(1999)\citenamefont{Sachdev}}]{Sachdev1999}
\bibinfo{author}{\bibfnamefont{S.}~\bibnamefont{Sachdev}}, 
\bibinfo{author}{\bibfnamefont{Quantum Phase Transitions (Cambridge Univ. Press, New York, 1999).}}


\bibitem[{\citenamefont{Custers et~al.}(2003)\citenamefont{Custers}}]{Custers2003}
\bibinfo{author}{\bibfnamefont{J.}~\bibnamefont{Custers}}, 
\bibinfo{author}{\bibfnamefont{P.}~\bibnamefont{Gegenwart}}, 
\bibinfo{author}{\bibfnamefont{H.}~\bibnamefont{Wilhelm}}, 
\bibinfo{author}{\bibfnamefont{K.}~\bibnamefont{Neumaier}}, 
\bibinfo{author}{\bibfnamefont{Y.}~\bibnamefont{Tokiwa}}, 
\bibinfo{author}{\bibfnamefont{O.}~\bibnamefont{Trovarelli}}, 
\bibinfo{author}{\bibfnamefont{C.}~\bibnamefont{Geibel}}, 
\bibinfo{author}{\bibfnamefont{F.}~\bibnamefont{Steglich}}, 
\bibinfo{author}{\bibfnamefont{C.}~\bibnamefont{Pepin}} and 
\bibinfo{author}{\bibfnamefont{P.}~\bibnamefont{Coleman}}, 
  \bibinfo{journal}{Nature} \textbf{\bibinfo{volume}{424}},
  \bibinfo{pages}{524} (\bibinfo{year}{2003}).

\bibitem[{\citenamefont{L¨ohneysen et~al.}(2007)\citenamefont{L¨ohneysen}}]{L¨ohneysen2007}
\bibinfo{author}{\bibfnamefont{H.v.}~\bibnamefont{Lohneysen}}, 
\bibinfo{author}{\bibfnamefont{A.}~\bibnamefont{Rosch}}, 
\bibinfo{author}{\bibfnamefont{M.}~\bibnamefont{Vojta}}  \bibnamefont{and}
\bibinfo{author}{\bibfnamefont{P.}~\bibnamefont{W¨olfle}}, 
  \bibinfo{journal}{Rev. Mod. Phys.} \textbf{\bibinfo{volume}{79}},
  \bibinfo{pages}{1015} (\bibinfo{year}{2007}).


\bibitem[{\citenamefont{Schr¨oder et~al.}(2000)\citenamefont{Schr¨oder}}]{Schr¨oder2000}
\bibinfo{author}{\bibfnamefont{A.}~\bibnamefont{Schr¨oder}}, 
\bibinfo{author}{\bibfnamefont{G.}~\bibnamefont{Aeppli}}, 
\bibinfo{author}{\bibfnamefont{R.}~\bibnamefont{Coldea}}, 
\bibinfo{author}{\bibfnamefont{M.}~\bibnamefont{Adams}}, 
\bibinfo{author}{\bibfnamefont{O.}~\bibnamefont{Stockert}}, 
\bibinfo{author}{\bibfnamefont{H.v.}~\bibnamefont{Lohneysen}}, 
\bibinfo{author}{\bibfnamefont{E.}~\bibnamefont{Bucher}}, 
\bibinfo{author}{\bibfnamefont{R.}~\bibnamefont{Ramazashvili}} and 
\bibinfo{author}{\bibfnamefont{P.}~\bibnamefont{Coleman}}, 
  \bibinfo{journal}{Nature} \textbf{\bibinfo{volume}{407}},
  \bibinfo{pages}{351} (\bibinfo{year}{2000}).
  



\bibitem[{\citenamefont{Gegenwart et~al.}(2006)\citenamefont{Gegenwart}}]{Gegenwart2006}
\bibinfo{author}{\bibfnamefont{P.}~\bibnamefont{Gegenwart}},
  \bibinfo{author}{\bibfnamefont{F.}~\bibnamefont{Weickert}},
  \bibinfo{author}{\bibfnamefont{M.}~\bibnamefont{Garst}},
  \bibinfo{author}{\bibfnamefont{R.S.}~\bibnamefont{Perry}} \bibnamefont{and}
  \bibinfo{author}{\bibfnamefont{Y.}~\bibnamefont{Maeno}}, 
  \bibinfo{journal}{Phys. Rev. Lett.} \textbf{\bibinfo{volume}{96}},
  \bibinfo{pages}{136402} (\bibinfo{year}{2006}).



\bibitem[{\citenamefont{Van der Marel et~al.}(2003)\citenamefont{Van der Marel}}]{VanderMarel2003}
\bibinfo{author}{\bibfnamefont{D.}~\bibnamefont{Van der Marel}}, 
\bibinfo{author}{\bibfnamefont{H.J.A.}~\bibnamefont{Molegraaf}}, 
\bibinfo{author}{\bibfnamefont{J.}~\bibnamefont{Zaanen}}, 
\bibinfo{author}{\bibfnamefont{Z.}~\bibnamefont{Nussinov}}, 
\bibinfo{author}{\bibfnamefont{F.}~\bibnamefont{Carbone}}, 
\bibinfo{author}{\bibfnamefont{A.}~\bibnamefont{Damascelli}}, 
\bibinfo{author}{\bibfnamefont{H.}~\bibnamefont{Eisaki}}, 
\bibinfo{author}{\bibfnamefont{M.}~\bibnamefont{Greven}}, 
\bibinfo{author}{\bibfnamefont{P.H.}~\bibnamefont{Kes}} and 
\bibinfo{author}{\bibfnamefont{M.}~\bibnamefont{Li}},
  \bibinfo{journal}{Nature} \textbf{\bibinfo{volume}{425}},
  \bibinfo{pages}{271} (\bibinfo{year}{2003}).


  

\bibitem[{\citenamefont{Georges et~al.}(1996)\citenamefont{Georges, Kotliar,
  Krauth, and Rozenberg}}]{Georges1996}
\bibinfo{author}{\bibfnamefont{A.}~\bibnamefont{Georges}},
  \bibinfo{author}{\bibfnamefont{G.}~\bibnamefont{Kotliar}},
  \bibinfo{author}{\bibfnamefont{W.}~\bibnamefont{Krauth}} \bibnamefont{and}
  \bibinfo{author}{\bibfnamefont{M.J.} \bibnamefont{Rozenberg}}, 
  \bibinfo{journal}{Rev. Mod. Phys.} \textbf{\bibinfo{volume}{68}},
  \bibinfo{pages}{13} (\bibinfo{year}{1996}); 
\bibinfo{author}{\bibfnamefont{P.}~\bibnamefont{Limelette}}, 
\bibinfo{author}{\bibfnamefont{A.}~\bibnamefont{Georges}}, 
\bibinfo{author}{\bibfnamefont{P.}~\bibnamefont{Wzietek}}, 
\bibinfo{author}{\bibfnamefont{D.}~\bibnamefont{Jerome}}, 
\bibinfo{author}{\bibfnamefont{P.}~\bibnamefont{Metcalf}} \bibnamefont{and}
\bibinfo{author}{\bibfnamefont{J.}~\bibnamefont{Honig}}, 
\bibinfo{journal}{Science} \textbf{\bibinfo{volume}{302}},
\bibinfo{pages}{89} (\bibinfo{year}{2003});
\bibinfo{author}{\bibfnamefont{P.}~\bibnamefont{Limelette}}, 
\bibinfo{author}{\bibfnamefont{P.}~\bibnamefont{Wzietek}}, 
\bibinfo{author}{\bibfnamefont{S.}~\bibnamefont{Florens}}, 
\bibinfo{author}{\bibfnamefont{A.}~\bibnamefont{Georges}}, 
\bibinfo{author}{\bibfnamefont{T.A.}~\bibnamefont{Costi}}, 
\bibinfo{author}{\bibfnamefont{C.}~\bibnamefont{Pasquier}},
\bibinfo{author}{\bibfnamefont{D.}~\bibnamefont{J\'erome}},
\bibinfo{author}{\bibfnamefont{C.}~\bibnamefont{M\'ezi\`ere}} \bibnamefont{and}
\bibinfo{author}{\bibfnamefont{P.}~\bibnamefont{Batail}}, 
\bibinfo{journal}{Phys. Rev. Lett.} \textbf{\bibinfo{volume}{91}},
\bibinfo{pages}{016401} (\bibinfo{year}{2003}).

  
  
  
  \bibitem[{\citenamefont{Terasaki et~al.}(1997)\citenamefont{Terasaki, Sasago and Uchinokura}}]{Terasaki1997}
\bibinfo{author}{\bibfnamefont{I.}~\bibnamefont{Terasaki}},
  \bibinfo{author}{\bibfnamefont{Y.}~\bibnamefont{Sasago}} \bibnamefont{and}
  \bibinfo{author}{\bibfnamefont{K.}~\bibnamefont{Uchinokura}}, 
  \bibinfo{journal}{Phys. Rev. B} \textbf{\bibinfo{volume}{56}},
  \bibinfo{pages}{12685 (R)} (\bibinfo{year}{1997}).
  


\bibitem[{\citenamefont{Limelette et~al.}(2006)\citenamefont{Limelette}}]{Limelette2006}
\bibinfo{author}{\bibfnamefont{P.}~\bibnamefont{Limelette}},
\bibinfo{author}{\bibfnamefont{S.}~\bibnamefont{H\'ebert}}, 
\bibinfo{author}{\bibfnamefont{V.}~\bibnamefont{Hardy}}, 
\bibinfo{author}{\bibfnamefont{R.}~\bibnamefont{Fr\'esard}}, 
\bibinfo{author}{\bibfnamefont{Ch.}~\bibnamefont{Simon}} \bibnamefont{and}
\bibinfo{author}{\bibfnamefont{A.}~\bibnamefont{Maignan}}, 
  \bibinfo{journal}{Phys. Rev. Lett.} \textbf{\bibinfo{volume}{97}},
  \bibinfo{pages}{046601} (\bibinfo{year}{2006}).


\bibitem[{\citenamefont{Bobroff et~al.}(2007)\citenamefont{Bobroff}}]{Bobroff2007}
\bibinfo{author}{\bibfnamefont{J.}~\bibnamefont{Bobroff}},
\bibinfo{author}{\bibfnamefont{S.}~\bibnamefont{H\'ebert}}, 
\bibinfo{author}{\bibfnamefont{G.}~\bibnamefont{Lang}}, 
\bibinfo{author}{\bibfnamefont{P.}~\bibnamefont{Mendels}}, 
\bibinfo{author}{\bibfnamefont{D.}~\bibnamefont{Pelloquin}} \bibnamefont{and}
\bibinfo{author}{\bibfnamefont{A.}~\bibnamefont{Maignan}}, 
  \bibinfo{journal}{Phys. Rev. B} \textbf{\bibinfo{volume}{76}},
  \bibinfo{pages}{100407 (R)} (\bibinfo{year}{2007}).



\bibitem[{\citenamefont{Takada et~al.}(2007)\citenamefont{Takada}}]{Takada2003}
\bibinfo{author}{\bibfnamefont{K.}~\bibnamefont{Takada}},
\bibinfo{author}{\bibfnamefont{H.}~\bibnamefont{Sakurai}}, 
\bibinfo{author}{\bibfnamefont{E.}~\bibnamefont{Takayama-Muromachi}}, 
\bibinfo{author}{\bibfnamefont{F.}~\bibnamefont{Izumi}}, 
\bibinfo{author}{\bibfnamefont{R.A.}~\bibnamefont{Dilanian}} \bibnamefont{and}
\bibinfo{author}{\bibfnamefont{T.}~\bibnamefont{Sasaki}}, 
  \bibinfo{journal}{Nature} \textbf{\bibinfo{volume}{422}},
  \bibinfo{pages}{53} (\bibinfo{year}{2003}).


\bibitem[{\citenamefont{Limelette et~al.}(2008)\citenamefont{Limelette}}]{Limelette2008}
\bibinfo{author}{\bibfnamefont{P.}~\bibnamefont{Limelette}},
\bibinfo{author}{\bibfnamefont{J.C.}~\bibnamefont{Soret}}, 
\bibinfo{author}{\bibfnamefont{H.}~\bibnamefont{Muguerra}} \bibnamefont{and}
\bibinfo{author}{\bibfnamefont{D.}~\bibnamefont{Grebille}}, 
  \bibinfo{journal}{Phys. Rev. B.} \textbf{\bibinfo{volume}{77}},
  \bibinfo{pages}{245123} (\bibinfo{year}{2008}); 
\bibinfo{author}{\bibfnamefont{P.}~\bibnamefont{Limelette}},
\bibinfo{author}{\bibfnamefont{S.}~\bibnamefont{H\'ebert}}, 
\bibinfo{author}{\bibfnamefont{H.}~\bibnamefont{Muguerra}}, 
\bibinfo{author}{\bibfnamefont{R.}~\bibnamefont{Fr\'esard}}, \bibnamefont{and}
\bibinfo{author}{\bibfnamefont{Ch.}~\bibnamefont{Simon}},
\bibinfo{journal}{Phys. Rev. B.} \textbf{\bibinfo{volume}{77}},
\bibinfo{pages}{235118} (\bibinfo{year}{2008}).

  
\bibitem[{\citenamefont{Li}(2004)\citenamefont{Li}}]{Li2004}
\bibinfo{author}{\bibfnamefont{S.Y.}~\bibnamefont{Li}}, 
\bibinfo{author}{\bibfnamefont{L.}~\bibnamefont{Taillefer}}, 
\bibinfo{author}{\bibfnamefont{D.G.}~\bibnamefont{Hawthorn}}, 
\bibinfo{author}{\bibfnamefont{M.A.}~\bibnamefont{Tanatar}}, 
\bibinfo{author}{\bibfnamefont{J.}~\bibnamefont{Paglione}}, 
\bibinfo{author}{\bibfnamefont{M.}~\bibnamefont{Sutherland}}, 
\bibinfo{author}{\bibfnamefont{R.W.}~\bibnamefont{Hill}}, 
\bibinfo{author}{\bibfnamefont{C.H.}~\bibnamefont{Wang}} and 
\bibinfo{author}{\bibfnamefont{X.H.}~\bibnamefont{Chen}}, 
  \bibinfo{journal}{Phys. Rev. Lett.} \textbf{\bibinfo{volume}{93}},
  \bibinfo{pages}{056401} (\bibinfo{year}{2004}).

  
\bibitem[{\citenamefont{Singh}(2003)\citenamefont{Singh}}]{Singh2007}
  \bibinfo{author}{\bibfnamefont{D.J.}~\bibnamefont{Singh}}, 
  \bibinfo{journal}{Phys. Rev. B} \textbf{\bibinfo{volume}{68}},
  \bibinfo{pages}{020503 (R)} (\bibinfo{year}{2003}).

 
\bibitem[{\citenamefont{Hervieu et~al.}(2003)\citenamefont{Hervieu}}]{Hervieu2003}
\bibinfo{author}{\bibfnamefont{M.}~\bibnamefont{Hervieu}},
\bibinfo{author}{\bibfnamefont{A.}~\bibnamefont{Maignan}}, 
\bibinfo{author}{\bibfnamefont{C.}~\bibnamefont{Michel}}, 
\bibinfo{author}{\bibfnamefont{V.}~\bibnamefont{Hardy}}, 
\bibinfo{author}{\bibfnamefont{N.}~\bibnamefont{Cr\'eon}} \bibnamefont{and}
\bibinfo{author}{\bibfnamefont{B.}~\bibnamefont{Raveau}}, 
  \bibinfo{journal}{Phys. Rev. B} \textbf{\bibinfo{volume}{67}},
  \bibinfo{pages}{045112} (\bibinfo{year}{2003}).



  \bibitem[{\citenamefont{Leligny et~al.}(1999)\citenamefont{Leligny}}]{Leligny1999}
\bibinfo{author}{\bibfnamefont{H.}~\bibnamefont{Leligny}},
\bibinfo{author}{\bibfnamefont{D.}~\bibnamefont{Grebille}}, 
\bibinfo{author}{\bibfnamefont{A.C.}~\bibnamefont{Masset}}, 
\bibinfo{author}{\bibfnamefont{M.}~\bibnamefont{Hervieu}}, 
\bibinfo{author}{\bibfnamefont{C.}~\bibnamefont{Michel}} \bibnamefont{and}
\bibinfo{author}{\bibfnamefont{B.}~\bibnamefont{Raveau}}, 
  \bibinfo{journal}{C.R. Acad. Sci., Ser. IIc: Chim} \textbf{\bibinfo{volume}{2}},
  \bibinfo{pages}{409} (\bibinfo{year}{1999}).
  

  
\bibitem[{\citenamefont{Gegenwart et~al.}(2005)\citenamefont{Gegenwart}}]{Gegenwart2005}
\bibinfo{author}{\bibfnamefont{P.}~\bibnamefont{Gegenwart}},
  \bibinfo{author}{\bibfnamefont{J.}~\bibnamefont{Custers}},
  \bibinfo{author}{\bibfnamefont{Y.}~\bibnamefont{Tokiwa}},
  \bibinfo{author}{\bibfnamefont{C.}~\bibnamefont{Geibel}} \bibnamefont{and}
  \bibinfo{author}{\bibfnamefont{F.}~\bibnamefont{Steglich}}, 
  \bibinfo{journal}{Phys. Rev. Lett.} \textbf{\bibinfo{volume}{94}},
  \bibinfo{pages}{076402} (\bibinfo{year}{2005}).


\bibitem[{\citenamefont{Hertz}(1976)\citenamefont{Hertz}}]{Hertz1976}
  \bibinfo{author}{\bibfnamefont{J.A.}~\bibnamefont{Hertz}}, 
  \bibinfo{journal}{Phys. Rev. B} \textbf{\bibinfo{volume}{14}},
  \bibinfo{pages}{1165-1184} (\bibinfo{year}{1976});
 \bibinfo{author}{\bibfnamefont{A.J.}~\bibnamefont{Millis}}, 
  \bibinfo{journal}{Phys. Rev. B} \textbf{\bibinfo{volume}{48}},
  \bibinfo{pages}{7183} (\bibinfo{year}{1993}).


\bibitem[{\citenamefont{Si et~al.}(2001)\citenamefont{Si}}]{Si2001}
  \bibinfo{author}{\bibfnamefont{Q.}~\bibnamefont{Si}},
  \bibinfo{author}{\bibfnamefont{S.}~\bibnamefont{Rabello}}, 
  \bibinfo{author}{\bibfnamefont{K.}~\bibnamefont{Ingersent}} \bibnamefont{and}
  \bibinfo{author}{\bibfnamefont{J.L.}~\bibnamefont{Smith}}, 
  \bibinfo{journal}{Nature} \textbf{\bibinfo{volume}{413}},
  \bibinfo{pages}{804} (\bibinfo{year}{2001}); 
  \bibinfo{author}{\bibfnamefont{Q.}~\bibnamefont{Si}},
  \bibinfo{author}{\bibfnamefont{S.}~\bibnamefont{Rabello}}, 
  \bibinfo{author}{\bibfnamefont{K.}~\bibnamefont{Ingersent}} \bibnamefont{and}
  \bibinfo{author}{\bibfnamefont{J.L.}~\bibnamefont{Smith}}, 
  \bibinfo{journal}{Phys. Rev. B} \textbf{\bibinfo{volume}{68}},
  \bibinfo{pages}{115103} (\bibinfo{year}{2003}).


  

  

\bibitem[{\citenamefont{Misawa}(2009)\citenamefont{Misawa}}]{Misawa2009}
  \bibinfo{author}{\bibfnamefont{T.}~\bibnamefont{Misawa}},
    \bibinfo{author}{\bibfnamefont{Y.}~\bibnamefont{Yamaji}} \bibnamefont{and}
  \bibinfo{author}{\bibfnamefont{M.}~\bibnamefont{Imada}}, 
  \bibinfo{journal}{J. Phys. Soc. Jpn:} \textbf{\bibinfo{volume}{78}},
  \bibinfo{pages}{084707} (\bibinfo{year}{2009}). 
 
\bibitem[{\citenamefont{Paul}(2001)\citenamefont{Paul}}]{Paul2001}
  \bibinfo{author}{\bibfnamefont{I.}~\bibnamefont{Paul}} \bibnamefont{and}
  \bibinfo{author}{\bibfnamefont{G.}~\bibnamefont{Kotliar}}, 
  \bibinfo{journal}{Phys. Rev. B} \textbf{\bibinfo{volume}{64}},
  \bibinfo{pages}{184414} (\bibinfo{year}{2001}).

\bibitem[{\citenamefont{Wang}(2003)\citenamefont{Wang}}]{Wang2003}
  \bibinfo{author}{\bibfnamefont{Y.}~\bibnamefont{Wang}},
    \bibinfo{author}{\bibfnamefont{N.S.}~\bibnamefont{Rogado}},
  \bibinfo{author}{\bibfnamefont{R.J.}~\bibnamefont{Cava}} \bibnamefont{and}
  \bibinfo{author}{\bibfnamefont{N.P.}~\bibnamefont{Ong}}, 
  \bibinfo{journal}{Nature} \textbf{\bibinfo{volume}{423}},
  \bibinfo{pages}{425} (\bibinfo{year}{2003}).



\end{thebibliography}
\end{document}